\documentclass[11pt]{article}
\usepackage{amsmath}
\usepackage{latexsym}
\usepackage{epsf}
\def\IR{\relax{\rm I\kern-.18em R}}
\def\I1{\relax{\rm 1\kern-.40em 1}}
\def\IZ{\relax{\rm Z\kern-.40em Z}}
\def\be{\begin{equation}}
\def\ee{\end{equation}}
\def\bq{\begin{eqnarray}}
\def\eq{\end{eqnarray}}
\begin{document}
\begin{flushright} LPTENS-08/27 \end{flushright}
\vskip 1cm
\begin{center}
{\bf \Large FOLLOWING THE PATH OF CHARM:\\
NEW PHYSICS AT THE LHC}
\vskip 1cm

{\Large JOHN ILIOPOULOS}
\vskip 1cm

Laboratoire de Physique Th\'eorique \\ de L'Ecole Normale Sup\'erieure \\
75231 Paris Cedex 05, France

\vskip 8cm

Talk presented at the ICTP on the occasion of the Dirac medal award ceremony
\vskip 1cm
Trieste, 27/03/08
\end{center}

\newpage

It is a great honour for me to speak on this occasion and I want to express my gratitude to the Abdus Salam International Centre for Theoretical Physics as well as the Selection Committee of the Dirac Medal. It is also a great pleasure to be here with Luciano Maiani and discuss the consequences of our common work with Sheldon Glashow on charmed particles \cite{GIM}. I will argue in this talk that the same kind of reasoning, which led us to predict the opening of a new chapter in hadron physics, may shed some light on the existence of new physics at the as yet unexplored energy scales of LHC.

The argument is based on the observation that precision measurements at a
given energy scale allow us to make predictions concerning the next energy
scale. It is remarkable that the origin of this observation can be traced back
to 1927, the two fundamental papers on the interaction of atoms with the
electromagnetic field written by Dirac, which are among the cornerstones of quantum
field theory. In the second of these papers \cite{Dirac} Dirac computes the scattering of light quanta by an atom $\gamma(k_1)+A_i \rightarrow \gamma(k_2)+A_f$, where $A_i$ and $A_f$ are the initial and final atomic states, respectively. He obtains the perturbation theory result:

\be
\label{Diracpert}
H_{fi}=\Sigma_j \frac{H^1_{fj}H^1_{ji}}{E_i-E_j}
\ee
where $H$ are the amplitudes. For the significance of the rhs, Dirac notes:
``...The scattered radiation thus appears as a result of the two processes $i
\rightarrow j$ and $j\rightarrow f$, one of which must be an absorption, the
other an emission, in neither of which the total proper energy is even
approximately conserved.'' This is the crux of the matter: In the calculation
of a transition amplitude we find contributions from states whose energy may
put them 
beyond our reach. The size of their contribution decreases with their energy,
see (\ref{Diracpert}), so, the highest the precision of our measurements, the
further away we can see. 

Let me illustrate the argument with two examples, one with a
non-renormalisable theory and one with a renormalisable one. A quantum field
theory, whether renormalisable or not, should be viewed as an effective theory
valid up to a given scale $\Lambda$. It makes no sense to assume a theory
for all energies, because we know already that at very high energies entirely
new physical phenomena appear (example: quantum gravity at the Planck
scale). The first  example is the Fermi four-fermion theory with a coupling
constant $G_F \sim 10^{-5}GeV^{-2}$. It is a non-renormalisable theory and, at
the $n$th order of perturbation, the $\Lambda$ dependence of a given quantity
$A$ is given by: 

\begin{equation}
A^{(n)}=C_0^{(n)} (G_F \Lambda^2)^n+ C_1^{(n)}G_F (G_F \Lambda^2)^{n-1}+
C_2^{(n)}G_F^2 (G_F \Lambda^2)^{n-2}+....
\label{Fermiexp}
\end{equation}
where the $C_i$'s are functions of the masses and external momenta, but their
dependence on $\Lambda$ is, at most, logarithmic. Perturbation theory breaks
down obviously when $A^{(n)} \sim A^{(n+1)}$ and this happens when $G_F
\Lambda^2 \sim 1$. This gives a scale of $\Lambda \sim 300$GeV as an upper
bound  for the validity of the Fermi theory. Indeed, we know today that at
100GeV the $W$ and $Z$ bosons change the structure of the theory. But, in
fact, we can do much better than that \cite{JOFSHAB}. Weak interactions violate some of the
conservation laws of strong interactions, such as parity and strangeness. The
absence of such violations in precision measurements will tell us that $G_F
\Lambda^2 \sim \epsilon$ with $\epsilon$ being the experimental precision. The
resulting limit depends on the value of the $C$ coefficient for the quantity
under consideration. In this particular case it turned out that, under the
assumption that the chiral symmetry of strong interactions is broken only by
terms transforming like the quark mass terms, the coefficient $C_0^{(n)}$ for parity and/or strangeness
violating amplitudes vanishes and no new limit is obtained \cite{BoucIlPrent}. However, the
second order coefficient $C_1^{(n)}$ contributes to flavour changing neutral
current transitions and the smallness of the $K_1-K_2$ mass difference, or the
$K^0_L \rightarrow \mu^+ +\mu^-$ decay amplitude, give a limit of $\Lambda
\sim 3$GeV before new physics should appear. The new physics in this case turned out to be the charmed
particles \cite{GIM}. We see in this example that the scale $\Lambda$ turned
out to be rather low and this is due to the non-renormalisable nature of the
effective theory which implies a power-law behaviour of the radiative
corrections on $\Lambda$.

The second example in which new physics has been discovered through its
effects in radiative corrections is the well-known ``discovery'' of the $t$
quark at LEP, before its actual production at Fermilab. The effective theory is now the Standard Model, which is renormalisable. In this case the dependence of the radiative corrections on the scale $\Lambda$ is, generically, logarithmic and the sensitivity of the low energy effective theory on the high scale is weak (there is an important exception to this rule for the Standard Model which we shall see presently). In spite of that,  the discovery
was made possible because of the special property of the Yukawa coupling
constants in the Standard Model to be proportional to the fermion
mass. Therefore, the effects of the top quark in the radiative corrections are
quadratic in $m_t$. The LEP precision measurements were able to extract a very
accurate prediction for the top mass. 

\begin{figure}
\centering
\epsfxsize=10cm
\epsffile{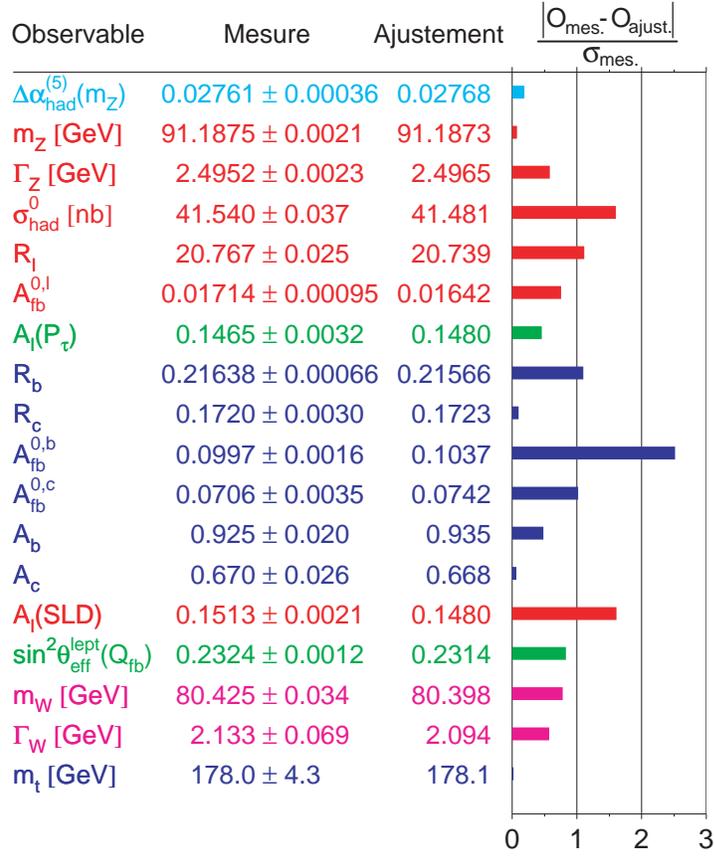}
\caption {Various physical quantities measured and computed.} \label{globfit}
\end{figure}

I claim that we are in a similar situation with the precision measurements of
the Standard Model. Our confidence in this model is amply justified on the basis of its ability to accurately describe the bulk of our present day 
data and, especially, of its enormous success in predicting new phenomena. All these spectacular successes are in fact successes of renormalised perturbation theory. 
Indeed what we have learnt was how to apply the methods which had been proven so powerful in quantum electrodynamics, 
to other elementary particle interactions. The remarkable quality of modern High Energy Physics experiments, mostly at LEP, 
but also elsewhere, has provided us with a large amount of data of unprecedented accuracy. All can be fit using the 
Standard Model with the Higgs mass as the only free parameter. Let me show some examples: Figure \ref{globfit} indicates the overall quality of such 
a fit. There are a couple of measurements which lay between 2 and 3 standard deviations away from the theoretical predictions, but
it is too early to say whether this is accidental, a manifestation of new physics, or the result of incorrectly combining incompatible
experiments.

\begin{figure}
\centering
\epsfxsize=6cm
\epsffile{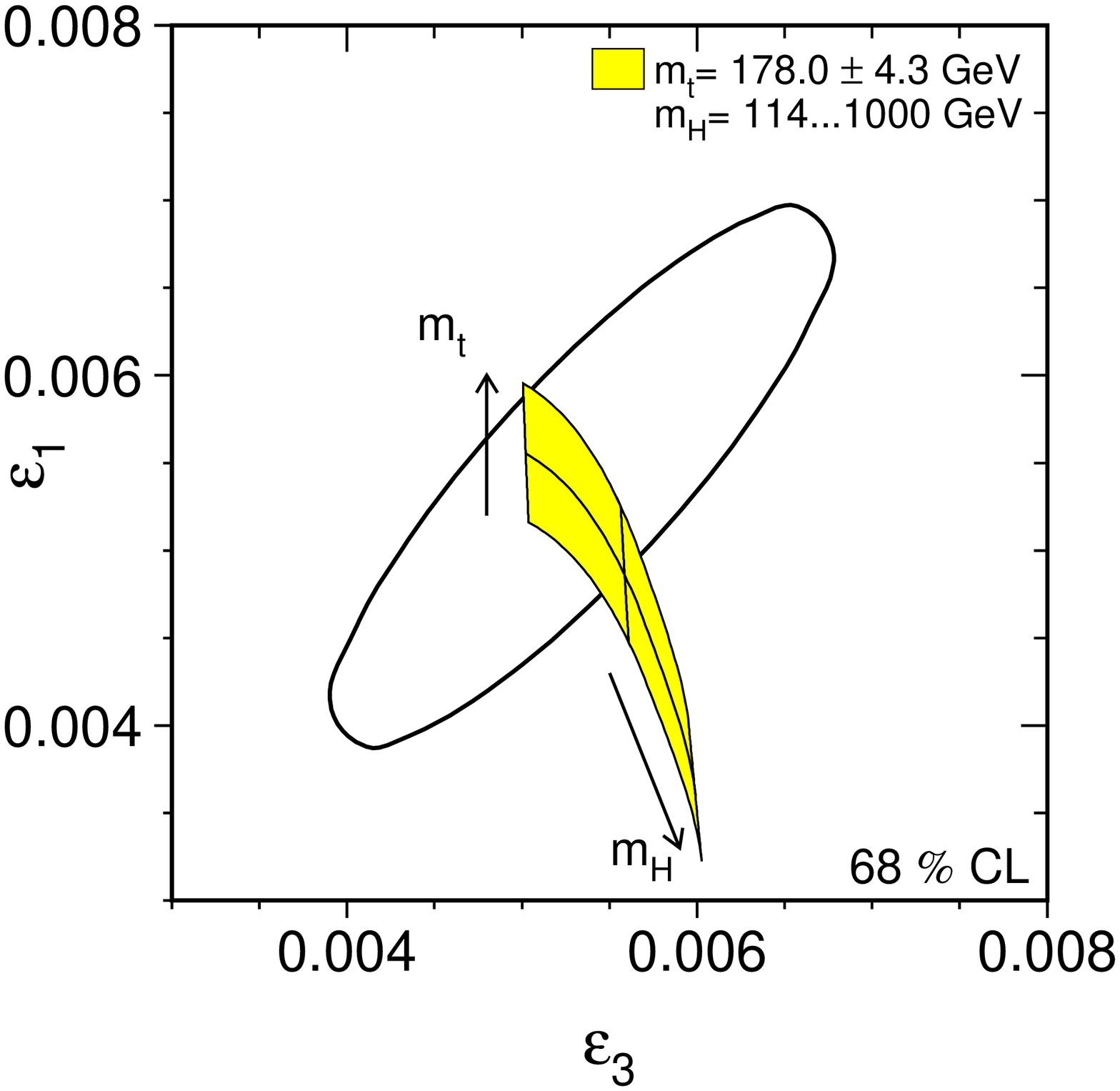}
\caption {The importance of the Standard Model radiative corrections. The
  arrows show how the prediction moves when we vary $m_t$ and $m_H$, in
  particular if we use the most recent lower value for  $m_t$.} \label{epsilon}
\end{figure}

Another impressive fit concerns the strong interaction effective coupling constant as a function of the momentum scale (Figure \ref{alphas}). This fit already shows the importance of taking into account the radiative corrections, since, in the tree approximation, $\alpha_s$ is obviously a constant. Similarly, Figure \ref{epsilon} shows the importance of the weak radiative corrections in the framework of the Standard Model. Because of the special Yukawa couplings, the dependence of these corrections on the fermion masses is quadratic, while it is only logarithmic in the Higgs mass. The $\epsilon$ parameters are designed to disentangle the two. The ones we use in Figure \ref{epsilon} are defined by:

\begin{figure}
\centering
\epsfxsize=6cm
\epsffile{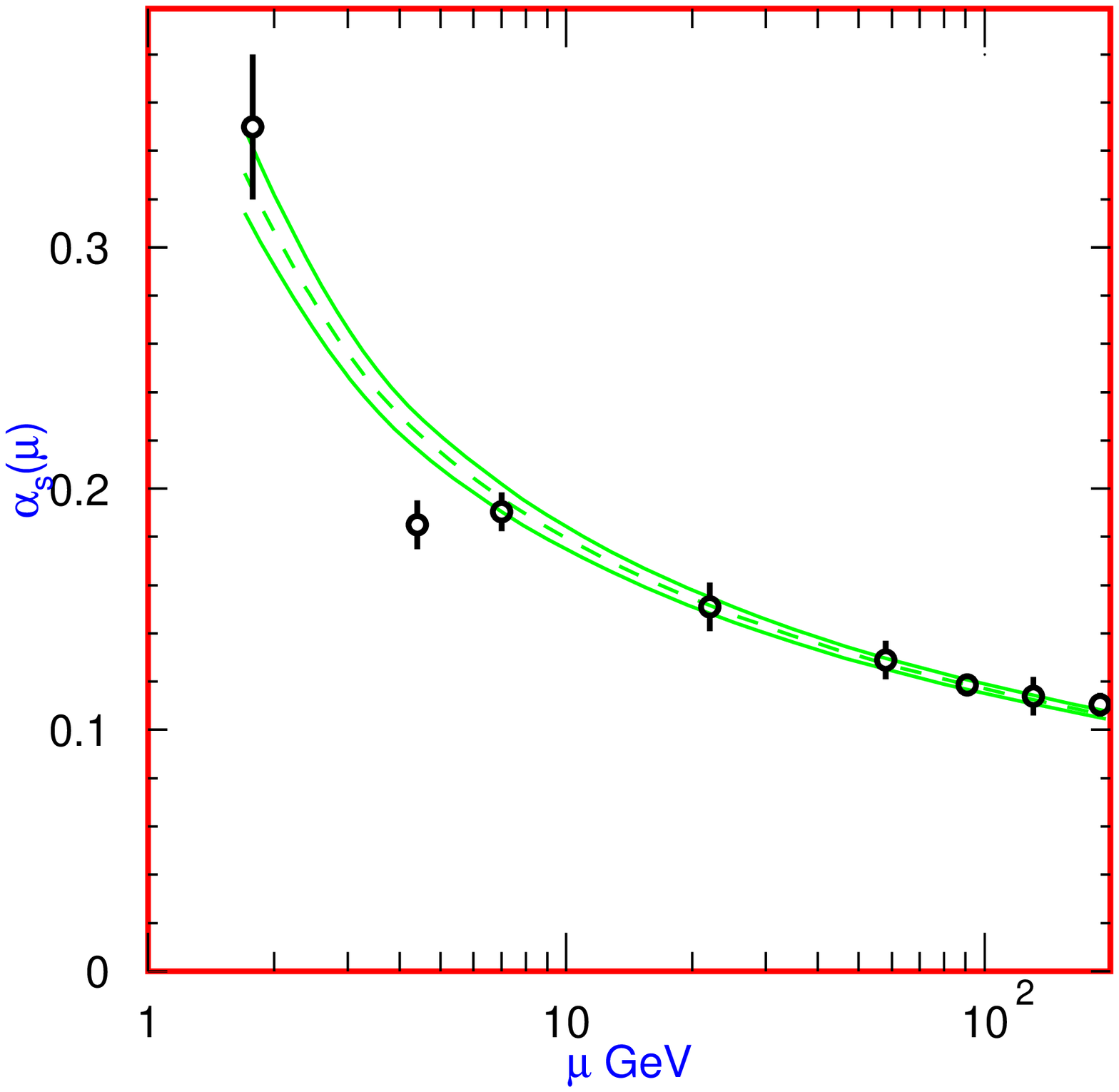}
\caption {The variation of $\alpha_s$ with the momentum scale. The renormalisation group prediction and the experimental points.} \label{alphas}
\end{figure}

\begin{equation}
\label{epspar1}
\epsilon_1=\frac{3G_Fm_t^2}{8\sqrt{2}\pi^2}-\frac{3G_Fm_W^2}{4\sqrt{2}\pi^2}\tan^2\theta_W \ln \frac{m_H}{m_Z}+...
\end{equation}

\begin{equation}
\label{epspar3}
\epsilon_3=\frac{G_Fm_W^2}{12\sqrt{2}\pi^2}\ln \frac{m_H}{m_Z}-\frac{G_Fm_W^2}{6\sqrt{2}\pi^2} \ln \frac{m_t}{m_Z}+...
\end{equation}
where the dots stand for subleading corrections. As you can see, the
$\epsilon$ s vanish in the absence of weak interaction radiative corrections,
in other words, $\epsilon_1=\epsilon_3=0$ are the values we get in the tree
approximation of the Standard Model but after having included the purely QED and QCD radiative corrections. We see clearly in Figure \ref{epsilon} that this point is excluded by the data. The latest values for these parameters are $\epsilon_1=5.4 \pm 1.0$ and $\epsilon_3=5.34 \pm 0.94$ \cite{Altar}. 

\begin{figure}
\centering
\epsfxsize=6cm
\epsffile{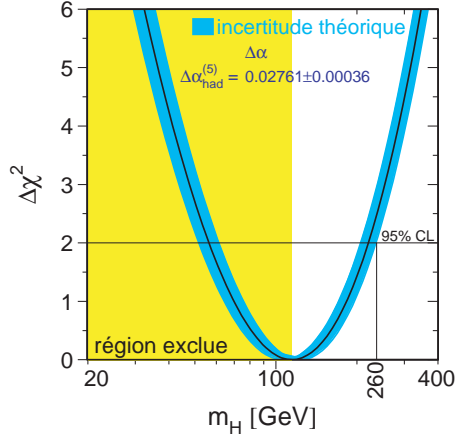}
\caption {The predicted values for the Standard Model Higgs mass using all available data. The shaded region is excluded by direct searches.} \label{limhiggs}
\end{figure}

Using all combined data we can extract the predicted values for the Standard
Model Higgs mass which are given in Figure \ref{limhiggs}. The data clearly
favour a low mass ($\leq 200$ GeV) Higgs, although, this prediction may be less
solid than what Figure \ref{limhiggs} seems to indicate. 

The main conclusion I want to draw from this comparison can be stated as follows:

{\it Looking at all the
data, from low energies to the Tevatron, we have learnt that perturbation
theory is remarkably successful, outside the specific regions where strong
interactions are important.} 

Let me explain this point better: At any given model with a coupling constant
$g$ we expect to have a weak coupling region $g\ll 1$, in which weak coupling
expansions, such as perturbation theory, are reliable, a strong coupling
region with $g\gg 1$, in which strong coupling expansions may be relevant, and
a more or less large gray region $g\sim 1$, in which no expansion is
applicable. The remarkable conclusion is that this gray
area appears to be extremely narrow. And this is achieved by an enlargement of
the area in which weak coupling expansion applies. The perturbation expansion is reliable,
not only for very small couplings, such as $\alpha_{em}\sim 1/137$, but also
for moderate QCD couplings $\alpha_s \sim 1/3$, as shown in Figure
\ref{alphas}. This is extremely important because without this property no
calculation would have been possible. If we had to wait until $\alpha_s$ drops
to values as low as $\alpha_{em}$ we could not use any available
accelerator. Uncalculable QCD backgrounds would have washed out any
signal. And this applies, not only to the Tevatron and LHC, but also to LEP.
We can illustrate this observation using a qualitative argument first introduced by F. Dyson. He noted that in a field theory like QED, the contribution of the $2n$-th order perturbation term to a physical amplitude $A^{(2n)}$ grows with $n$ roughly as\footnote{The estimation is only heuristic. It is based on a rough counting of the number of diagrams and assumes that they all contribute equally and have the same sign, neither of which is exact. The estimation can be improved but the result remains the same.} 

\be
A^{(2n)} \sim \alpha^{n} (2n-1)!!
\ee
where $\alpha$ is (the square of) the coupling constant. Again, perturbation theory will break down when $A^{(2n)}\sim A^{(2n+2)}$ which gives

\be
2n+1 \sim \alpha^{-1}
\ee

This leaves a comfortable margin for QED but leaves totally unexplained the successes of QCD at moderate energies. 

\begin{figure}
\centering
\epsfxsize=12cm
\epsffile{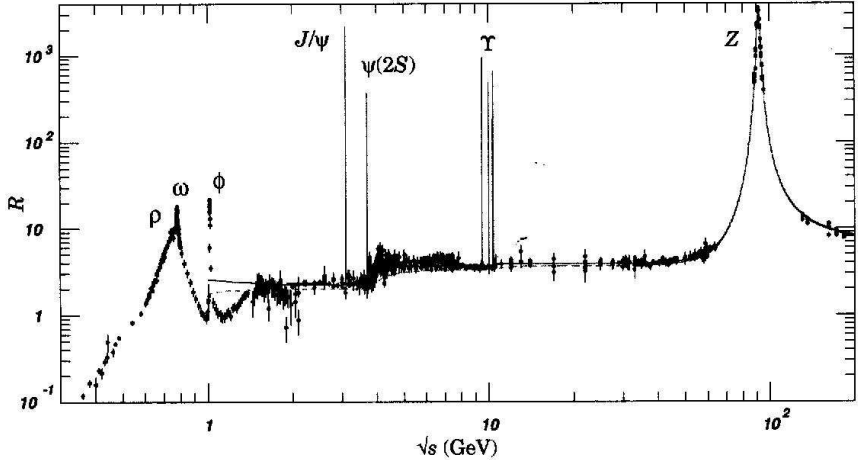}
\caption {The ratio $R$ of $e^+ +e^-$ total cross section to
hadrons normalised to that of $e^+ +e^- \rightarrow \mu^+ +\mu^-$ as a
function of the centre-of-mass energy.} \label{R}
\end{figure}

A
global view of the weak and strong coupling regions is given in Figure \ref{R}
which shows the $R$-ratio, {\it i.e.} the $e^+ +e^-$ total cross section to
hadrons normalised to that of $e^+ +e^- \rightarrow \mu^+ +\mu^-$ as a
function of the centre-of-mass energy. The lowest order perturbation value for
this ratio is a constant, equal to $\Sigma Q_i^2$, the sum of the squares of
the quark charges accessible at this energy. We see clearly in this Figure the
areas of applicability of perturbation theory: At very low energies, below 1
GeV, we are in the strong coupling regime characterised by resonance
production. The strong interaction effective coupling constant becomes of
order one (we can extrapolate from Figure \ref{alphas}), and perturbation
breaks down. However, as soon as we go slightly above 1 GeV, $R$ settles to
a constant value and it remains such except for very narrow regions when new
thresholds open. In these regions the cross section is again dominated by
resonances and perturbation breaks down. But these areas are extremely well
localised and  threshold effects do not spread outside these small regions. 

In this talk I want to exploit this observational fact and argue that the available
precision tests of the Standard Model allow us to claim with confidence that
new physics will be unravelled at the LHC, although
we have no unique answer on the nature of this new physics.  The argument assumes the
validity of perturbation theory and it will fail if the latter fails. But, as
we just saw, perturbation theory breaks down only when strong interactions
become important. But new strong interactions do imply new physics.

The key is again the Higgs boson. As we explained above, the data favour a low mass Higgs. However,
the opposite cannot be excluded, first because it depends on the subset of the
data one is looking at\footnote{This prediction is, in fact, an average
  between a much lower value, around 50 GeV, given by the data from leptonic
  asymmetries, and a much higher one, of 400 GeV, obtained from the hadronic
  asymmetries. Although the difference sounds dramatic, the two are still
  mutually consistent at the level of 2-3 standard deviations.}, and, second,
because the analysis is done taking the minimal Standard Model.

Given this result, let us see what, if any, are the theoretical constraints. The Standard Model Higgs mass is given, at the classical level, by $m_H^2=2\lambda v^2$, with $v$ the vacuum expectation value of the Higgs field. $v$ is fixed by the value of the Fermi coupling constant $G_F/\sqrt 2 =1/(2v^2)$ which implies $v\approx$246 GeV. Therefore, any constraints will come from the allowed values of $\lambda$. A first set of such constraints is given by the classical requirement:

\begin{equation}
\label{clconstr}
1>\lambda >0 ~~~~\Rightarrow ~~~~ m_H<400-500 GeV
\end{equation}
 
The lower limit for $\lambda$ comes from the classical stability of the
theory. If $\lambda$ is negative the Higgs potential is unbounded from below
and there is no ground state. The upper limit comes from the requirement of
keeping the theory in the weak coupling regime. If $\lambda \geq 1$ the Higgs
sector of the theory becomes strongly interacting and we expect to see
plenty of resonances and bound states rather than a single elementary
particle. 

Going to higher orders is straightforward, using the renormalisation group
equations. The running of the effective mass is determined by that of
$\lambda$. Keeping only the dominant terms and assuming $t=\log (v^2/\mu^2)$ is
small ($\mu \sim v$), we find 

\begin{figure}
\centering
\epsfxsize=12cm
\epsffile{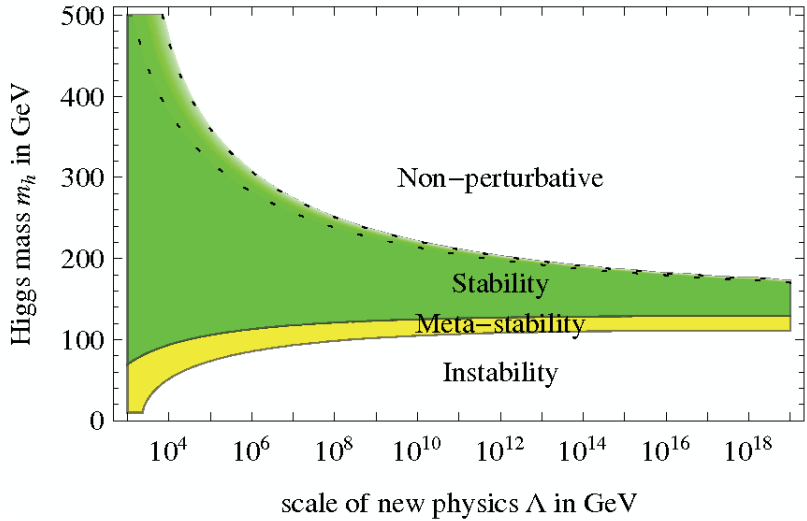}
\caption {Bounds on the Higgs mass.} \label{bounds}
\end{figure}

\begin{equation}
\label{quconstr1}
\frac{d\lambda}{dt}=\frac{3}{4\pi^2}[\lambda^2+3\lambda h_t^2-9h_t^4+...]
\end{equation}
where $h_t$ is the coupling of the Higgs boson to the top quark. The dots
stand for less important terms, such as the other Yukawa couplings to the
fermions and the couplings with the gauge bosons. This equation is correct as
long as all couplings remain smaller than one, so that perturbation theory is
valid, and no new physics beyond the standard model becomes important. Now we
can repeat the argument
on the upper
and lower bounds for $\lambda$ but this time taking into account the full
scale dependence $\lambda (\mu)$. We thus obtain for the Higgs mass an upper bound given by
the requirement of weak coupling regime ($\lambda (\mu)<1$) all the way up to
the scale $\mu$, and a lower bound
by the requirement of vacuum stability ($\lambda (\mu)>0$), again up to $\mu$. Obviously, the
bounds will be more stringent the larger the assumed value of $\mu$. Figure
\ref{bounds} gives the allowed region for the Higgs mass as a function of the
scale for scales up to the Planck mass. We see that for small $\mu \sim 1$TeV,
the limits are, essentially, those of the tree approximation equation (\ref{clconstr}),
while for $\mu \sim M_P$ we obtain only a narrow window of allowed masses
130GeV$<m_H<$200GeV, remarkably similar to the experimental results.

This analysis gives the first conclusion: If perturbation theory remains valid, in other words, if we have no new strong interactions, there exists at least one, relatively light, Higgs boson:
\vskip 0.5cm

{\it Conclusion 1: The absence of a light Higgs boson implies New Physics.}   
\vskip 0.5cm

Here ``heavy Higgs'' is not clearly distinguished from ``no-Higgs'', because a very heavy
Higgs, above 1 TeV, is not expected to appear as an elementary particle. As we
explained above, this will be accompanied by new strong interactions. A
particular version of this possibility is the ``Technicolor'' model,  which
assumes the existence of a new type of fermions with strong interactions at
the multi-hundred-GeV scale. 
The role of the Higgs is played by a fermion-antifermion bound state. ``New
Physics''  is precisely the discovery of a completely new sector of
elementary particles. Other strongly interacting models can and have been constructed. The general conclusion
 is that a heavy Higgs always implies new forces whose effects are
expected to be visible at the LHC.\footnote{We can build specific models in which the
  effects are well hidden and pushed above the LHC discovery
  potential, at least with the kind of accuracy one can hope to achieve in a hadron machine. In this case one would need very high precision measurements,
  probably with a multi-TeV $e^+-e^-$ collider.} 

The possibility which seems to be favoured by the data is the presence of a ``light'' Higgs particle. In this case new strong interactions are not needed and, therefore, we can assume that perturbation theory remains valid. But then we are faced with a new problem. The Standard Model is a renormalisable theory and the dependence on the high energy scale is expected to be only logarithmic. This is almost true, but with one notable exception: The radiative corrections to the Higgs mass are quadratic in whichever scale $\Lambda$ we are using. The technical reason is that $m_H$ is the only parameter of the Standard Model which requires, by power counting, a quadratically divergent counterterm. The gauge bosons require no mass counterterm at all because they are protected by gauge invariance and the fermions need only a logarithmic one. The physical reason is that, if we put a fermion mass to zero we increase the symmetry of the model because now we can perform chiral transformations on this fermion field. Therefore the massless theory will require no counterterm, so the one needed for the massive theory will be proportional to the fermion mass and not the cut-off. In contrast, putting $m_H=0$ does not increase the symmetry of the model.\footnote{At the classical level, the Standard Model with a massless Higgs does acquire a new symmetry, namely scale invariance, but this symmetry is always broken for the quantum theory and offers no protection against the appearance of quadratic counterterms.} As a result the effective mass of the Higgs boson will be given by

\begin{equation}
\label{mhiggseff}
(m_H^2)_{eff}=m_H^2+C\alpha_{eff}\Lambda^2
\end{equation}   
where $C$ is a calculable numerical coefficient of order one and $\alpha_{eff}$ some effective coupling constant. In practice it is dominated by the large coupling to the top quark. The moral of the story is that the Higgs particle cannot remain light unless there is a precise mechanism to cancel this quadratic dependence on the high scale. This is a particular aspect of a general problem called ``scale hierarchy''. The only known mechanism which reconciles a light Higgs and a high value of the scale $\Lambda$ with the validity of perturbation theory is supersymmetry. In this case the Higgs mass is protected against the quadratic corrections of eq. (\ref{mhiggseff}) because it behaves like the mass of the companion fermion which, as we just said, receives only logarithmic corrections. It is closest in spirit to the charm mechanism, in the sense that a heavy effective cut-off is made compatible with the low energy data by the presence of new particles. 
The alternative is to have a low value of $\Lambda$, {\it i.e.} new physics, at a low scale. The models with large compact extra dimensions enter into this category. This brings us to our second conclusion:
\vskip 0.5cm

{\it Conclusion 2: A light Higgs boson is unstable without new physics.} 
\vskip 0.5cm

Both conclusions are good news for LHC. But the time for speculations is
coming to an end. The LHC is coming. Never before a new experimental facility
had such a rich discovery potential and never before was it loaded with so high expectations.

\end{document}